\documentclass[fleqn,usenatbib]{mnras}

\newcommand{\g}{$\gamma$-ray}
\newcommand{\lat}{\textit{Fermi}-LAT}

\usepackage{newtxtext,newtxmath}
\usepackage[T1]{fontenc}

\DeclareRobustCommand{\VAN}[3]{#2}
\let\VANthebibliography\thebibliography
\def\thebibliography{\DeclareRobustCommand{\VAN}[3]{##3}\VANthebibliography}

\usepackage{graphicx}	
\usepackage{amsmath}	

\title[Berkeley 87]{Diffuse gamma-ray emission in the vicinity of open cluster Berkeley 87}

\author[Ou \& Yang]{
Ziwei Ou,$^{1}$\thanks{E-mail: ziwei@sjtu.edu.cn}
and Xiaolong Yang$^{2,3}$\thanks{E-mail: yangxl@shao.ac.cn}
\\
$^{1}$Tsung-Dao Lee Institute, Shanghai Jiao Tong University, Shanghai, 200240, China\\
$^{2}$Shanghai Astronomical Observatory, Chinese Academy of Sciences, Shanghai 200030, China\\
$^{3}$Shanghai Key Laboratory of Space Navigation and Positioning Techniques, Shanghai 200030, China
}

\date{Accepted XXX. Received YYY; in original form ZZZ}

\pubyear{\the\year{}}

\begin{document}
\label{firstpage}
\pagerange{\pageref{firstpage}--\pageref{lastpage}}
\maketitle

\begin{abstract}
Young massive stellar clusters have been recognized as a significant population of cosmic ray accelerators and gamma-ray sources. We report the detection of extended gamma‑ray emission toward the young massive star cluster Berkeley 87 using 18 years of \textit{Fermi}-LAT data. The emission has an angular extension of 0.36 degree and a photon index of 2.68. To investigate the ambient gas target, we combine CO, EBHIS and \textit{Planck} data. The total gas mass within the region is $\sim 21.61 \times 10^2 M_{\odot}$, with an average proton density of $\sim 368\ \rm cm^{-3}$. The hadronic scenario is favored given the dense gas and the cluster’s strong stellar winds. The kinetic energy injection rate from stellar winds in Berkeley 87 exceeds $1.0 \times 10^{36}\ \rm erg \ s^{-1}$. Considering a CR acceleration efficiency $10^{-2.0}$ to $10^{-2.5}$, the cluster can supply sufficient energy to power the observed gamma-ray emission. A nearby pulsar, PSR~J2021+3651, associated with the Dragonfly pulsar wind nebula and the 4HWC source 4HWC~J2021$+$3650, lies in the same field of view. We also discuss the maximum accelerated proton energy and electron energy density of this pulsar.
\end{abstract}




\section{Introduction}
\label{sec:intro}

During the 17 years for its operation, \lat\ has found numerous high-energy \g\ sources \citep{Abdollahi2020}. The \g\ morphological and spectral information would assist us in understanding the nature of \lat\ sources. Observations from the \lat\ on supernova remnants (SNRs) \citep{Araya2024}, pulsar wind nebulae (PWNe) \citep{Tibaldo2018,Principe2020}, pulsar halos \citep{DiMauro2019,Abdollahi2024} and young massive stellar clusters (YMSCs) found extended \g\ emissions around them \citep{Peron2024b}. 

YMSCs host lots of massive stars including OB and Wolf‑Rayet (WR) stars, which drive powerful stellar winds that sustain strong shocks capable of accelerating particles to relativistic energies \citep{Voelk1982,Maurin2016,Aharonian2019,Peron2024a}. In OB stars, the wind–interstellar medium (ISM) interaction forms a termination shock where particles can be re‑accelerated, while WR stars exhibit much higher mass‑loss rates and wind speeds, often leading to wind–wind collision zones in binaries. These shocks are efficient sites of diffusive shock acceleration (DSA). Thus, \g\ emission from YMSCs arises from both leptonic and hadronic processes, which can be detected by \lat\ . In WR binaries, the wind‑collision region is particularly favourable for hadronic emission \citep{Benaglia2003,Bednarek2005}. The injection of particles occurs on short timescales inside the collision zone, while the overall activity is modulated by the orbital motion and the long‑term ($\sim$ Myr) evolution of the massive stars.

When two stellar winds collide as in massive binaries, a double‑shock structure with a contact discontinuity forms. The relative wind speed can reach several thousand $\rm km\ s^{-1}$, producing highly turbulent magnetic fields amplified to mG levels. This environment is an efficient particle accelerator for protons \citep{Torres2004,Reimer2006,Farnier2011}. The \g\ emissions can be produced by hadronic processes because of the high target gas density in the shocked wind region \citep{Bednarek2014,Bykov2020}. Such emission have been observed in systems like WR 140 and $\eta$ Carinae \citep{Abdo2010,Marti-Devesa2021}. In contrast, an isolated star with a non‑colliding wind expands into the ISM, forming only a single termination shock with much lower speed and lower ambient gas density. Particle acceleration is less efficient, limiting the maximum energy. The emission is much harder to detect with current \g\ telescopes \citep{Aliu2008}. These fundamental differences highlight that colliding‑wind binaries are promising \g\ sources \citep{Werner2013,Pshirkov2016}.

The collective action of stellar winds from massive stars may carve out large cavities known as stellar‑wind bubbles \citep{Meyer2024}. The wind sweeps up the ISM, creating a multi‑layered structure, with a freely expanding inner wind, a hot shocked plasma bubble, and a dense shell of swept‑up material \citep{Ackermann2011}. On larger scales, winds from an entire star cluster merge to form giant superbubbles \citep{Bykov2014}, which can break out of the galactic disk and drive kpc‑scale superwinds. These wind‑driven structures affect CR acceleration and transport. The forward shocks bounding the bubbles and the termination shocks of superbubbles are efficient sites of DSA, with model estimates showing that the wind kinetic energy can be converted into relativistic particles. The bubble interior also provides a target for CRs interactions, producing \g\ emission.


Berkeley 87 is one of the most massive star clusters in our Galaxy. This star cluster has an age of $6.5 \pm 0.4$ Myr and a total stellar mass of about 1551 $M_{\odot}$ \citep{Raudeliunas2026}. It hosts at least 15 OB stars and a WR star WR 142 \citep{Sokal2010}. The interaction of the strong ($\sim$ 5200 km/s) stellar wind from the WR 142, with the cluster stars and molecular clouds appears to produce dissipative shock waves \citep{Polcaro1991}. The missive stars with strong stellar winds embed in a very dense interstellar medium and molecular clouds. Molecular clouds are also observed in this region, whose velocity dispersion reveals a hint of the cloud collision. The distance to Berkeley 87 is estimated to be $1.67 \pm 0.02$ kpc \citep{Fuente2021}. By applying the acceleration process at the shocks arising in the winds of WR stars, the high-energy radiation processes of Berkeley 87 were explored by \cite{Bednarek2007}. During the \textit{EGRET} era, some stellar clusters have been found in the region of the \textit{EGRET} unidentified sources despite of the relatively large error boxes. The HEGRA Collaboration performs observations on Berkeley 87 but find no TeV \g\ \citep{Tluczykont2001}. In the third \textit{EGRET} catalog, 3EG~J2021$+$4716 and 3EG~J2016$+$3657 are suspected to be spatial related to Berkeley 87 \citep{Hartman1999}. It was suggested that Berkeley 87 correlates with at least one unassociated \lat\ source \citep{Peron2024b}. The TeV source MGRO~J2019$+$37 locate around the position of Berkeley 87 \citep{Abdo2007}.

In this paper, we investigate the particle propagation of Berkeley 87 by analyzing \lat\ data. We discuss the diffused \g\ emissions around this stellar cluster.

\section{Data Analysis}

\subsection{Spatial analysis}

We selected LAT data towards Berkeley 87 for a period of approximately 18 yr. For the analysis, we used the standard LAT analysis software package Fermitools \textit{v11r5p3} and Fermipy \textit{v1.2.0}. The region of interest (ROI) was selected to be a $15^{\circ} \times 15^{\circ}$ square centred on the position of Berkeley 87 (R.A.= 305.396$^{\circ}$, Dec.=37.4206$^{\circ}$). In order to reduce the effect of the Earth albedo background, we excluded the time intervals when the parts of the ROI were observed at zenith angles $> 90^{\circ}$. This energy cut reduces the contribution of bright, nearby pulsars. The P8\_R3\_v6 version of the post-launch instrument response functions (IRFs) was used and both the front and back converted photons were selected.

\begin{figure*}
    \centering
    \includegraphics[width=0.8\linewidth]{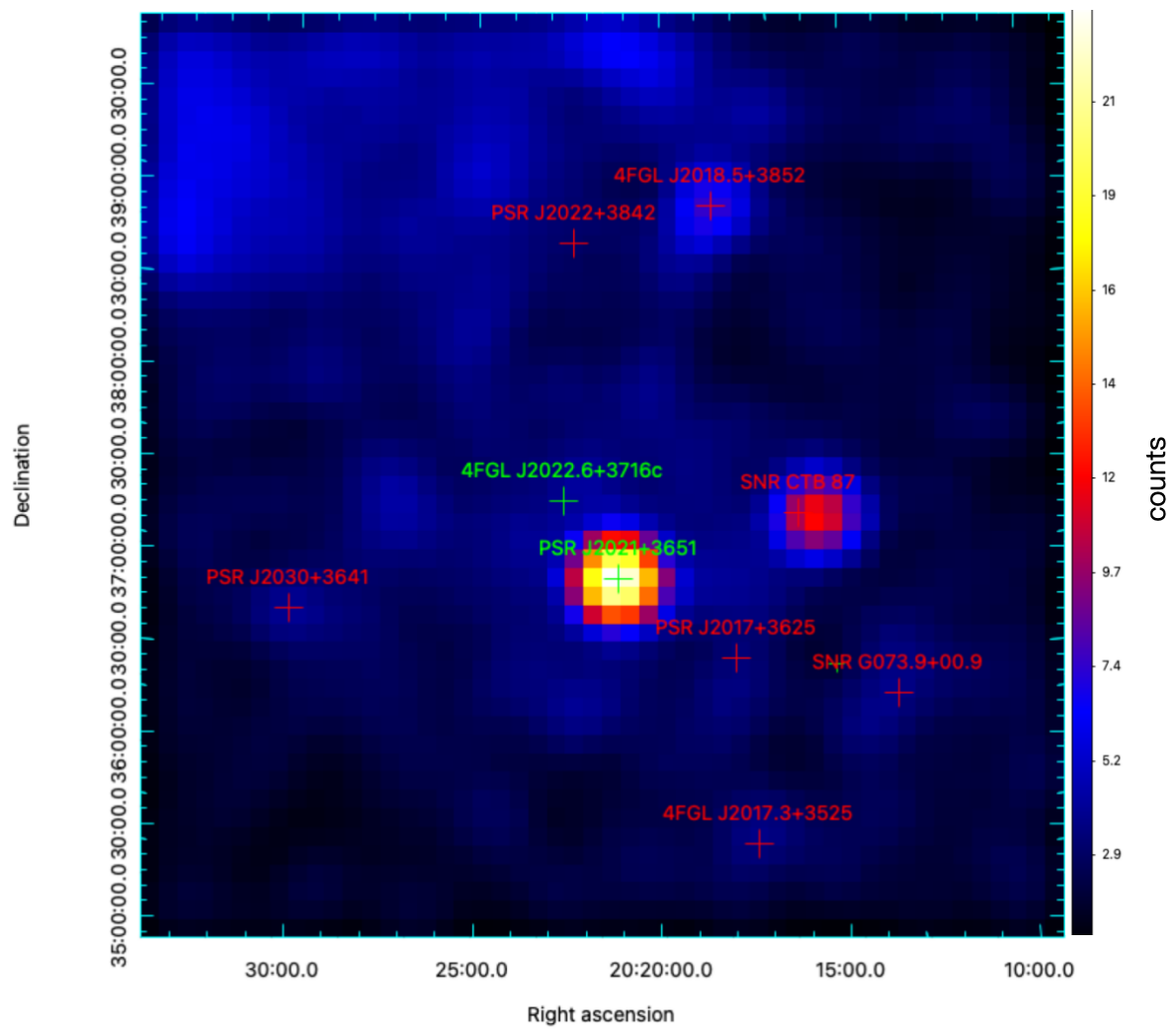}
    \caption{\g\ count map above 10 GeV in the $5^{\circ} \times 5^{\circ}$ region around Berkeley 87. The color bar provides the counts values of each pixel. 4FGL~J2022.6$+$3716 and PSR~J2021$+$3651 are shown in green cross, while other 4FGL sources are shown in red cross.}
    \label{fig:cmap}
\end{figure*}

For the spatial analysis, given the crowded nature of the region and to avoid systematic errors due to a poorer angular resolution at low energies, we selected only the photons with energies exceeding 10 GeV. The \g\ count map above 10 GeV in the $5^{\circ} \times 5^{\circ}$ region around Berkeley 87 is shown on Figure~\ref{fig:cmap}. A \g\ pulsar PSR~J2021$+$3651 shows significant \g\ emissions above 10 GeV. For further investigating the emissions around Berkeley 87, the on-pulse of PSR~J2021$+$3651 need to removed. Detail of pulsar gating analysis can be found in Section~\ref{sub:gating}. We performed a binned likelihood analysis via the tool \texttt{gtlike}. The point sources listed in the 4FGL-DR4 source catalog are adopted in the analysis.

\begin{figure*}
    \centering
    \includegraphics[width=0.8\linewidth]{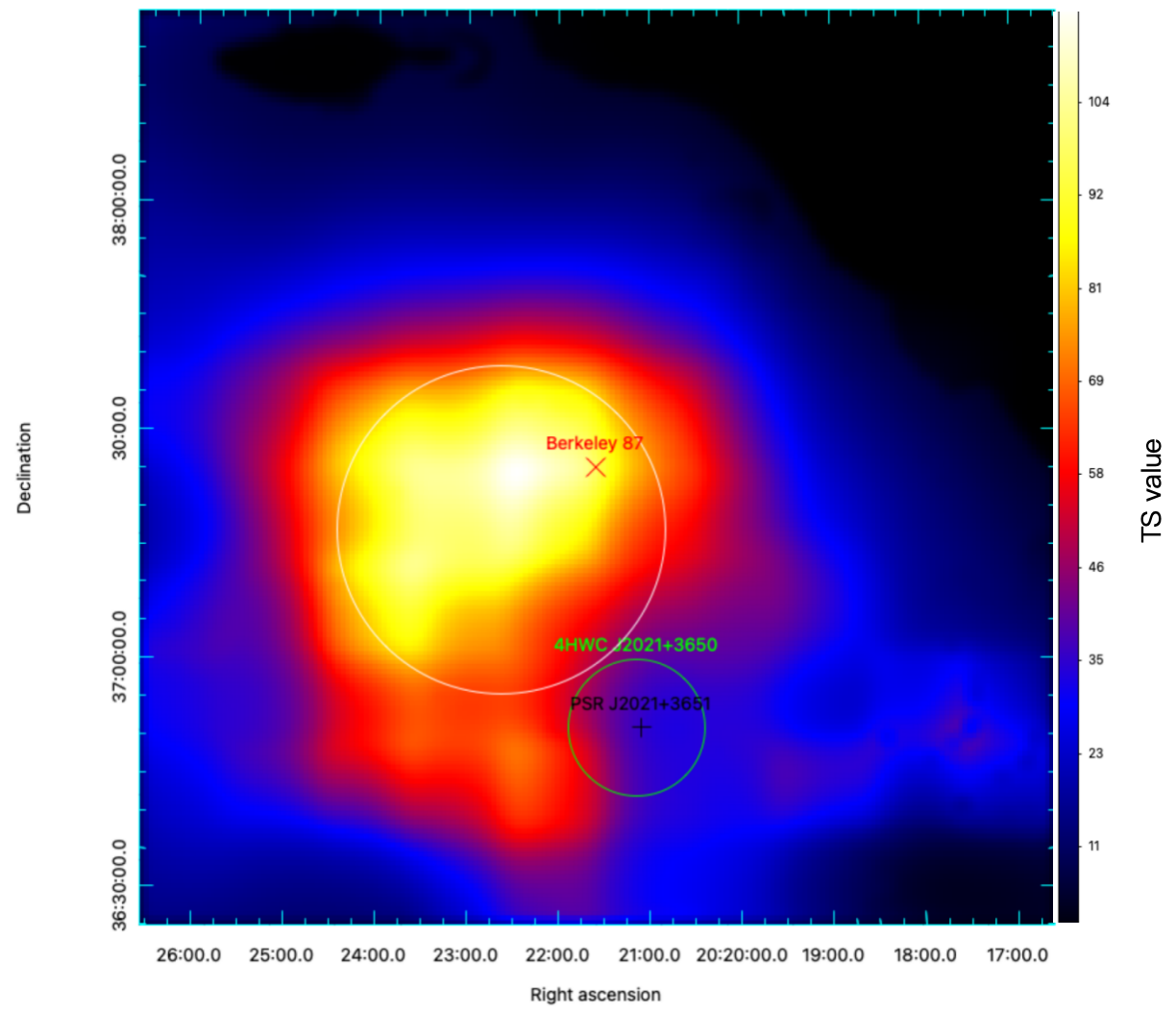}
    \caption{$2^{\circ} \times 2^{\circ}$ test statistic (TS) map. The color bar provides the TS values of each pixel. The estimated radius of 4FGL~J2022.6$+$3716 is shown in white circle. The central position of Berkeley 87 is marked as x. The pulsar PSR~J2021$+$3651 \citep{Smith2023} and radius of 4HWC~J2021$+$3650 \citep{Alfaro2026} are shown also.}
    \label{fig:ts-map}
\end{figure*}

The test statistic (TS) is adopted to estimate the significance of \g\ sources. It is defined by TS = 2 (ln $L_1$ - ln $L_0$), where $L_1$ and $L_0$ represent maximum likelihood values for background with target source and without target source. We removed 4FGL~J2022.6$+$3716 from the model. Figure~\ref{fig:ts-map} shows the TS map centered at position of Berkeley 87. We performed a binned likelihood analysis to obtain the value log(L) and the Akaike information criterion (AIC). The AIC is defined by AIC = -2log(L) + 2k, where k is the number of free parameters in the model. Corresponding results are provided in Table~\ref{tab:fit-model}.

A source extension analysis is executed for each source by performing a likelihood ratio test with respect to the point source with a best-fit model for extension. 

\begin{equation}\label{eq:ext}
    TS_{\rm ext} = -2 \left( \mathrm{ln}\left(L_{\rm PS}\right) -  \mathrm{ln}\left(L_{\rm ext}\right) \right)
\end{equation}

For obtaining TS$_{\rm ext}$, we adopt radial Gaussian and radial disk to model the spatial extension of Berkeley~87. The extension fitting find the radius as $0.36 \pm 0.07 ^{\circ}$ and $0.29 \pm 0.08 ^{\circ}$ for 2D radial Gaussian and 2D radial disk, respectively (Table~\ref{tab:fit-model}). The TS$_{\rm ext}$ for these two models are 27.6 and 26.1. To investigate whether gas around Berkeley 87 and \g\ emission correlation, we considered a spatial template combining molecular hydrogen (H$_2$) and ionized hydrogen (HII). Detail of gas analysis can be found in Section \ref{sec:gas}. For the H$_2$ + HII template, TS$_{\rm ext}$ = 20.4 is obtained. Therefore, we adopt a radial Gaussian model with $R_{68} = 0.36^{\circ}$ for our analysis.

\begin{table*}
    \centering
    \begin{tabular}{l|c|c|c|c|c|c}
    \hline
        Model & -log(L) & TS & TS$_{\rm ext}$ & $R_{68}$ & d.o.f. & $\Delta$AIC \\\hline
        Point Source & -21698360.03 & 103.1 & - & - & 56 & - \\
        Gaussian & - 21698265.78 & 130.2 & 27.6 & 0.36 &  57 & -186.5 \\
        Disk & -21698292.08 & 129.5 & 26.1 & 0.29 & 57 & -133.9 \\
        H$_2$ + HII & -21698309.83 & 122.7 & 20.4 & 0.18 & 58 & -96.4 \\
    \hline
    \end{tabular}
    \caption{Analysis results of different spatial models.}
    \label{tab:fit-model}
\end{table*}

\subsection{Spectral analysis}

For the spectral analysis we applied \textit{gtlike} in the energy range 300 MeV to 500 GeV and modelled the spectrum of radial Gaussian. We neglected the emission below 300 MeV to avoid the effect from the Galactic diffuse background. We adopt a power law (PL) spectrum model for the analysis. The PL model is given by: 

\begin{equation}
    \frac{dN}{dE} = N_0 \left( \frac{E}{E_0} \right)^{-\gamma}
\end{equation}
where $N_0$, $E_0$ and $\gamma$ are normalization, scale and photon index, respectively.

\begin{figure}
    \centering
    \includegraphics[width=0.48\textwidth]{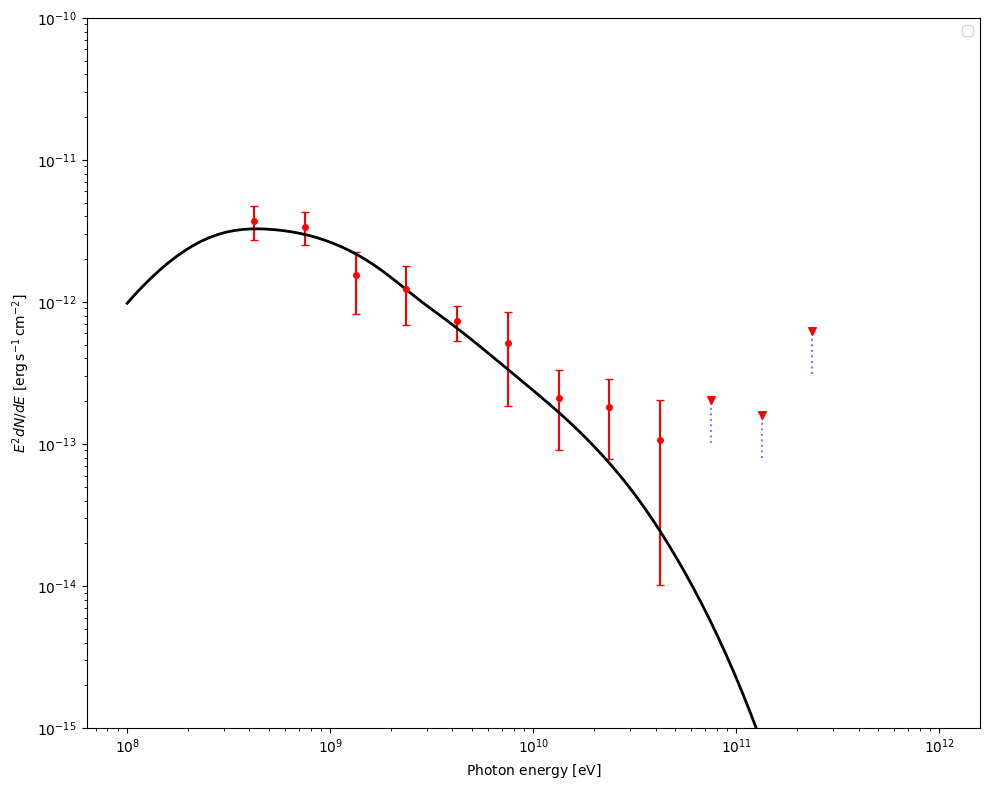}
    \caption{\g\ spectrum of Berkeley 87 with \lat\ data. The line shows the hadronic model.}
    \label{fig:sed}
\end{figure}

We divided the full energy range into 12 logarithmically spaced bands and applied \textit{gtlike} to each of these bands, in this process the index was fixed to 2. We derived the spectral energy distribution (SED), which are shown in Figure~\ref{fig:sed}. When the TS value of spectral point is less than 9, the upper limit is calculated at the 95\% confidence level. The best-fit photon index is $2.68 \pm 0.34$.

\subsection{Pulsar gating analysis of PSR~J2021$+$3651}
\label{sub:gating}

\begin{figure}
    \centering
    \includegraphics[width=0.95\linewidth]{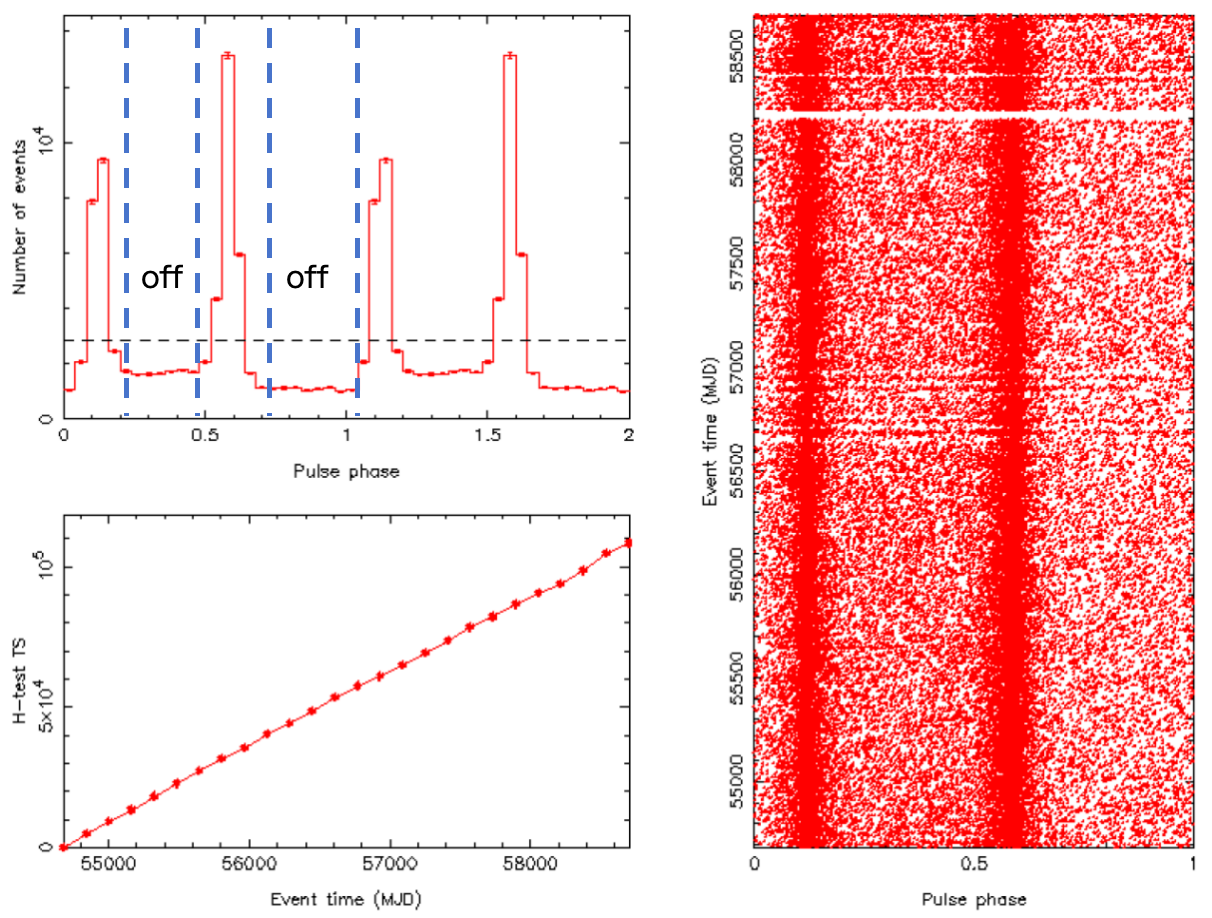}
    \caption{PSR~J2021$+$3651 timing results from \textit{Tempo2} with the \textit{Fermi} plug-in. Top-left panel: phase histogram of the analyzed \lat\ data. Two full rotational phase are shown here. Bottom-left panel: H-test significance as a function of time. Right panel: pulse phase for each \g\ event vs. time.}
    \label{fig:profile}
\end{figure}

Photons from PSR~J2021$+$3651 within a radius of 0.6$^{\circ}$ and a minimum energy of 300 MeV were selected, which maximized the H-test statistic. We adopt the pulsar ephemeris from 3PC catalog \citep{Smith2023}. \textit{Tempo2} with \textit{Fermi} plug-in were used to produce \g\ pulse profile of SR J2021$+$3651 \citep{Ray2011}. We were able to obtain Fourier template profiles for the whole time span, generate the times of arrival (TOAs), and obtain timing solutions by fitting the TOAs with frequency derivatives, which are shown in Figure~\ref{fig:profile}. The off-peak interval in our analysis is then defined to be at $\phi$ = (0-0.08) \& (0.16-0.51) \& (0.65-1).

We conduct the analysis by searching radius of PSR~J2021$+$3651 in the off-pulse phase as we do for Berkeley 87. A radial Gaussian model is used to obtain TS$_{\rm ext}$. The best-fit extension values are determined by performing a likelihood profile scan over the 68\% containment and fitting for the extension which maximizes the model. TS$_{\rm ext}$ for the off-pulse component is 2.1, which imply a point-like source for off-pulse. 

To validate the spectral model, we test for spectral curvature that reveals deviations from a PL spectrum for each source via likelihood ratio test. For an exponential cutoff power-law (PLSC) model, it gives: TS$_{\rm PLSC}$ = -2 $\left( \mathrm{ln}\left(L_{\rm PL}\right) - \mathrm{ln}\left(L_{\rm PLSC}\right) \right)$. The curvature test shows a marginal value TS$_{\rm PLSC} = 247$ for off-pulse. Therefore, a PLSC model should be considered to describe off-pulse, which means the off-pulse emission originates from pulsar magnetosphere. 

\section{Gas around Berkeley 87}
\label{sec:gas}

To study the interaction between stellar winds and the surrounding gas, we investigate distribution of molecular hydrogen H$_2$, neutral atomic hydrogen HI, and ionized hydrogen HII in the vicinity of Berkeley 87.

\begin{figure*}
    \centering
    \includegraphics[width=0.45\linewidth]{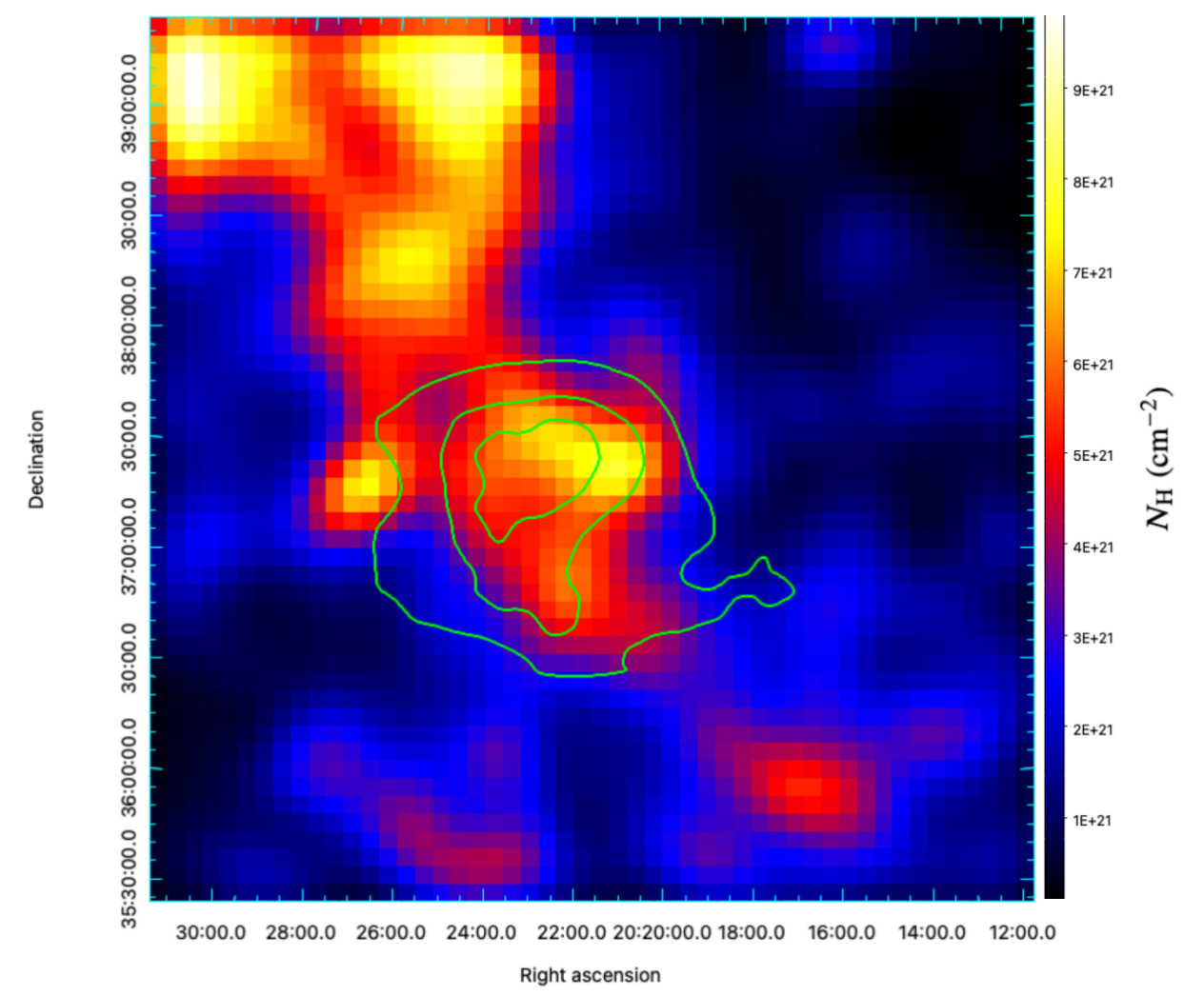}
    \includegraphics[width=0.45\linewidth]{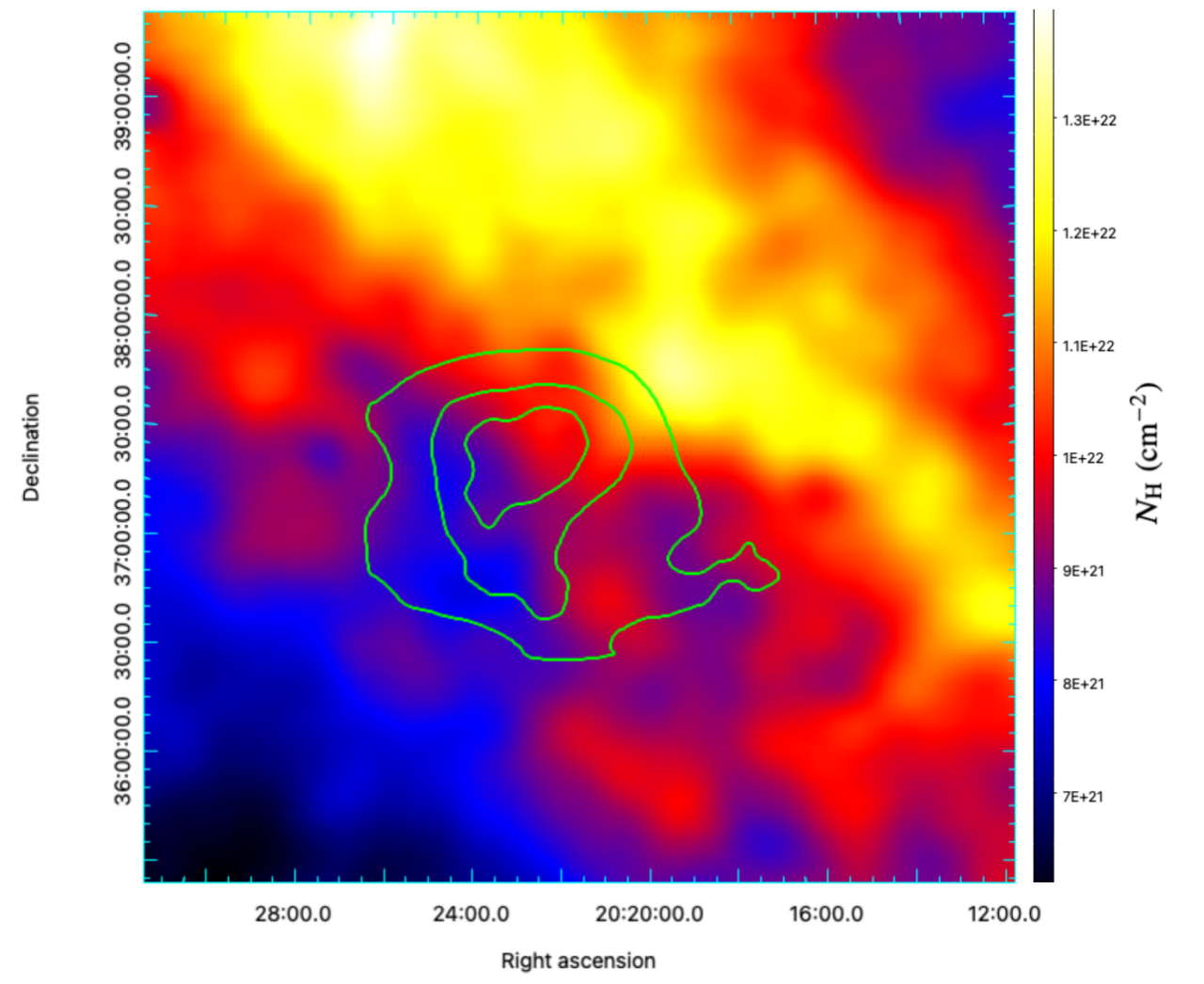}
    \includegraphics[width=0.45\linewidth]{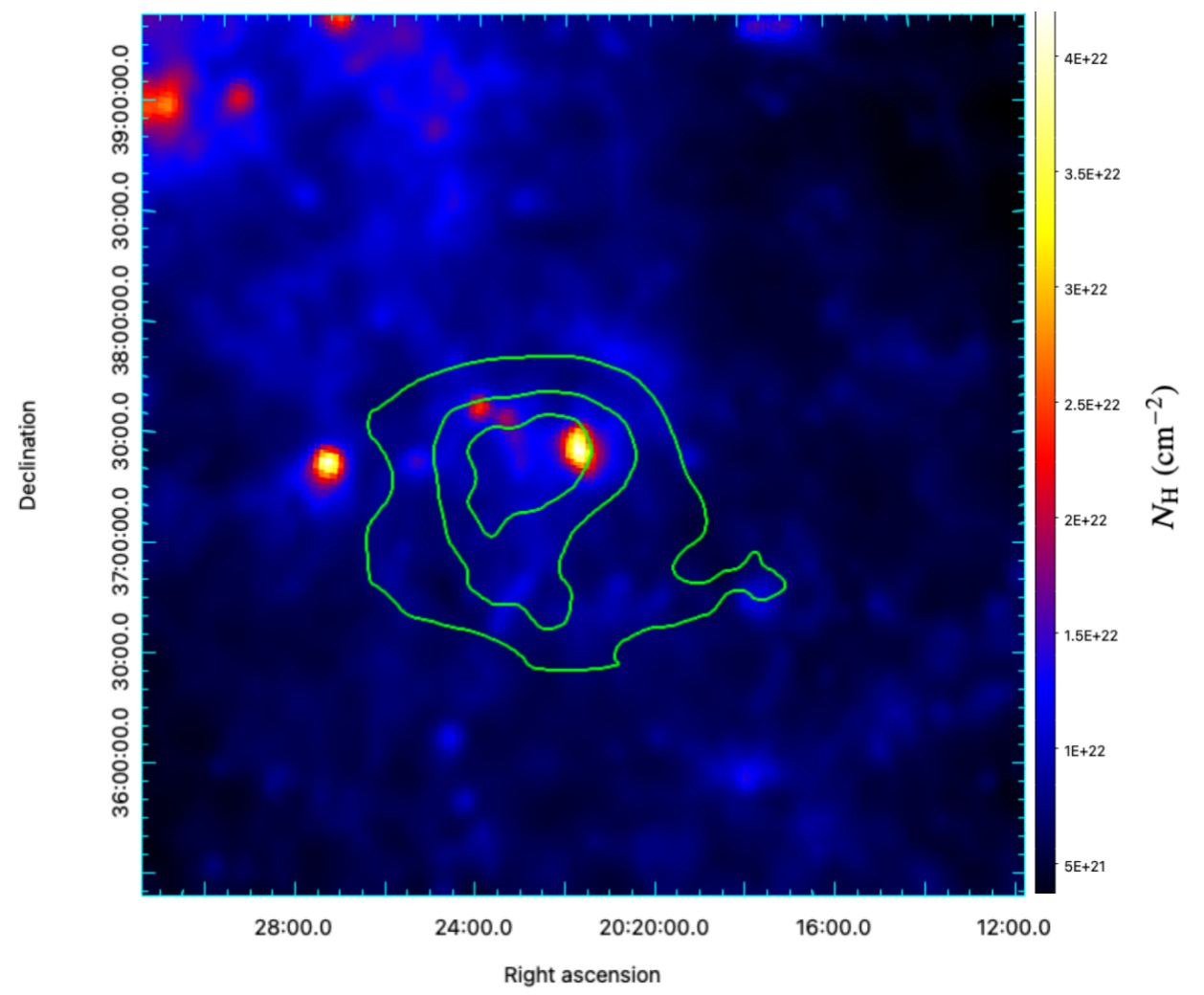}
    \caption{Gas column densities in different phase. The color bars represent the column density $N_{\rm H}$ ($\rm cm^{-2}$) with calculation method in Section~\ref{sec:gas}. The top-left panel gives the $\rm H_2$ column density derived from CO data. The top-right panel provides the map of H I column density obtained from EBHIS survey. The bottom panel shows the H II column density derived from \textit{Planck} 353 GHz map assuming an effective density of electron $n_{\rm e} = 10\ \rm cm^{-3}$. The \g\ emission is presented as green contours.}
    \label{fig:gas}
\end{figure*}

To investigate the H$_2$ distribution toward Berkeley~87, we use CO data from \cite{Dame2001}. The $^{13}$CO (J = 1-0) line profile of the molecular cloud may reflect the kinematic activity of the gas distribution. For Berkeley 87, the velocity range of the CO is considered to be -20 to 20 $\rm km\ s^{-1}$ \citep{Roy2011}. The column density of H$_2$ can be defined by $N_{\rm H_2} = X_{\rm CO} \times W_{\rm CO}$. The conversion factor $X_{\rm CO}$ is set to $2.0 \times 10^{20}\ \rm cm^{-2}\ K^{-1}\ km^{-1}\ s$ \citep{Dame2001,Bolatto2013}.

The CO line intensity $W_{\rm CO}$ (in unit of $\rm K\ km\ s^{-1}$) was calculated over the velocity range specified above. Figure~\ref{fig:gas} (top-left) shows the column density of H$_2$. The total mass of the molecular cloud can be calculated with 

\begin{equation}
    M = \mu m_{\rm H} d^2 \Omega_{\rm px} X_{\rm CO} \sum_{\rm px} W_{\rm CO}
\end{equation}

where $\mu$ reflects the mean molecular cloud, $m_{\rm H}$ is the mass of the H, and $\Omega_{\rm px}$ is the solid angle for each pixel. 

The HI data are obtained from the Effelsberg-Bonn HI Survey (EBHIS) \citep{Winkel2016}. This survey aims to obtain northern-hemisphere part of HI data. The number density $N_{\rm HI}$ can be calculated as:

\begin{equation}
   N_{\rm HI} = -1.38 \times 10^{18}\ T_{\rm s} \int d\nu \ln (1- \frac{T_{\rm B}}{T_{\rm s} - T_{\rm bg}})
\end{equation}

where $T_{\rm B}$ and $T_{\rm bg}$ are brightness temperatures of the HI emission, and the cosmic microwave background, respectively. We adopted $T_{\rm bg} = 2.66\ \rm K$ for the calculation. For $T_{\rm B} > T_{\rm s} - 5\ \rm K$, $T_{\rm B}$ can be approximated as $T_{\rm s} - 5\ \rm K$, and $T_{\rm s}$ is taken to be 150 K. Corresponding column density map of HI is presented in Figure~\ref{fig:gas} (top-right).

The H II column density can be derived using the \textit{Planck} free-free map \citep{Planck2016}. To do this, the emission measure must be converted into free–free intensity. The conversion factor for this transformation is provided in \cite{Finkbeiner2003}. The column density can then be calculated from the free-free emission: 

\begin{equation} 
    N_{\rm H II}  = 1.2 \times 10^{15} \ (\frac{T_{\rm e}}{1\ \rm K})^{0.35} (\frac{\nu}{1 \ \rm GHz})^{0.1} (\frac{n_{\rm e}}{1\ \rm  cm^{-3}})^{-1} (\frac{I_{\nu}}{1\ \rm Jy\ sr^{-1}})
\end{equation}

$N_{\rm H II}$ is in unit of $\rm cm^{-2}$. In such case, the frequency $\nu = 353\ \rm GHz$, electron temperature $T_{\rm e} = 8000\ \rm K$, and the effective electron density $n_{\rm e} = 10\ \rm cm^{-3}$ are adopted in the calculation. $I_{\nu} = 46.04\ \rm Jy\ sr^{-1}$ is adopted corresponding to $\nu = 353\ \rm GHz$ \citep{Finkbeiner2003}. Figure~\ref{fig:gas} (bottom) gives the derived HII column density.

The total gas mass of about $21.62 \times 10^2\ \rm M_{\odot}$ and the average proton density of $368 \ \rm cm^{-3}$ provide a sufficient target mass for \g\ production via pion decay (Table~\ref{tab:gas-mass}).

\section{Origin of diffuse $\gamma$-ray emissions around Berkeley 87}

\subsection{CRs injection and diffusion from Berkeley 87}

YMSCs are emerging as prominent Galactic sources of cosmic rays CRs, potentially dominating the CR population up to PeV energies. The intense collective stellar winds of O and WR stars power strong terminal and wind-wind collision shocks that efficiently accelerate particles via DSA. The escaped CRs from Berkeley 87 may interact with the dense ambient gas and radiation fields, producing extended GeV \g\ emissions, which extend beyond the core of Berkeley 87 as the CRs propagate outward. This scenario has been observationally verified in several YMSCs, such as Westerlund 1 \citep{Abramowski2012} Westerlund 2 \citep{Yang2018} and Berkeley~59 \citep{Ou2026}, where the detected \g\ morphology and spectrum support a scenario of efficient CR acceleration within the cluster and subsequent diffusion into the ISM, thus significantly contributing to the Galactic CR sea \citep{Aharonian2019,Aharonian2020}.

Considering the mass-loss rate $\dot{M}$ and wind velocity $v_{\rm w}$, the kinetic energy injection rate from the wind of the a single star is defined by $L_{\rm w} = \frac{1}{2} \dot{M} v_{\rm w}^2$. Typical values are given to $\dot{M} = 10^{-7}\ M_{\odot}\ \rm yr^{-1}$ and $v_{\rm w} = 2000\ \rm km\ s^{-1}$, which result in $L_{\rm w} \approx 1.0 \times 10^{35}\ \rm erg \ s^{-1}$. Given that Berkeley 87 hosts at least 15 OB stars and a WR star, the total kinetic energy injection rate should be $> 1.0 \times 10^{36}\ \rm erg \ s^{-1}$. The energy injection rate into CRs by stellar winds is defined by $\dot{E}_{\rm CR} = \eta_{\rm CR} L_{\rm w}$, where $\eta_{\rm CR}$ is the fraction of the stellar wind kinetic energy that goes into accelerating CRs. \cite{Batzofin2026} consider the $10^{-2.0}$ and $10^{-2.5}$ as $\eta_{\rm CR}$ values, which represented Kraichnan diffusion and Kolmogorov diffusion. Thus, $\dot{E}_{\rm CR} \sim 10^{34} \ \rm erg \ s^{-1}$ in this scenario.

The efficient injection of relativistic particles from YMSCs into the ambient ISM is primarily governed by diffusive transport \citep{Morlino2021,Menchiari2025}. The extended and spatially smooth morphology of the \g\ emission observed around clusters, which can be found in several typical YMSCs. The particles, having been accelerated at cluster wind termination shocks and inside wind-wind collision zones, scatter off turbulent magnetic fluctuations in the ISM. This diffusive behaviour naturally produces a radially declining surface brightness profile in \g\ \citep{Yang2018,Bhadra2022}. The effective diffusion coefficient inside the cavity and near the boundary of a YMSC can be significantly suppressed compared to the canonical Galactic average value of approximately $10^{28}\ \rm cm^2\ s^{-1}$ at 1 TeV. This suppression, often by a factor of 10–100, results from strong magnetic field amplification driven by the copious supply of CRs from the cluster, leading to a highly turbulent, scattering-dominated medium.

\subsection{Hadronic and leptonic scenarios}

\cite{Bednarek2007} suggests that the most likely description of the radiation processes in Berkeley 87 is achieved in the hybrid leptonic–hadronic model in which leptons are responsible for the observed X-ray and GeV \g\  emission and hadrons are responsible for the TeV \g\ emission.

If Berkeley~87 can accelerate protons to high energies, then these protons could illuminate the molecular clouds around them via proton–proton interaction. We assume a power-law spectrum for the parent proton distribution: $f_{\rm p}(E) = A_{\rm p} (E/E_0)^{-\alpha_{\rm p}}$. We use \texttt{NAIMA} package to fit the SED \citep{Zabalza2015}. Considering the distance of about 1.67 kpc, the best-fit proton index $\alpha_{\rm p} = 2.82 \pm 0.05$ and the total proton energy $W_{\rm p} = 5.2 \times 10^{49}\ \rm erg$ above 5 GeV are obtained. As a result, Berkeley 87 is powerful to accelerate enough CRs to account for the \g\ emissions under hadronic scenario.

We also tested the potential leptonic scenario of Berkeley~87. It reflects the \g\ generated via the inverse Compton (IC) scattering of electrons off the seed photons around the source. The OB stars lie in this stellar cluster provide potential radiation fields. We considered the CMB radiation field, optical to UV radiation field from the star light as photon field, and the dust infrared radiation field \cite{Kafexhiu2014}. In addition, nonthermal bremsstrahlung radiation resulting from the interaction between relativistic particles and thermal particle population may contribute \g\ emissions \citep{Baring1999}. The cooling timescale of bremsstrahlung can be derived by \citep{Blumenthal1970}:
\begin{equation}
    t_{\rm br} \simeq 4 \times 10^4 (\frac{10^3 \ \rm cm^{-3}}{n_{\rm gas}})\ \rm yr
\end{equation}

Thus, we obtain $t_{\rm br} \approx 1.09 \times 10^5 \ \rm yr$. We calculated the IC and bremsstrahlung spectrum using \texttt{NAIMA} as we did for hadronic scenario. By assuming a power-law distribution of the relativistic electrons: $f_{\rm e}(E) = A_{\rm e} (E/E_0)^{-\alpha_{\rm e}}$, the best-fit electron index of electron distribution and the total energy above 5 GeV can be obtained as $\alpha_{\rm e} = 1.72 \pm 0.04$ and $W_{\rm e} = 1.04 \times 10^{49}\ \rm erg$, respectively. Therefore, the leptonic origin of Berkeley~87 can not be ruled out.

\begin{table}
    \centering
    \begin{tabular}{c|c|c}
    \hline
    Tracer & Mass ($10^2\ M_{\odot}$) & Density ($\rm cm^{-3}$)\\\hline 
        H$_2$ & 3.03 & 52\\
        HI & 10.82 & 184\\
        HII & 7.77 & 132 \\
        Total & 21.62 & 368\\ \hline
    \end{tabular}
    \caption{Gas mass and number density derived from various tracer.}
    \label{tab:gas-mass}
\end{table}

\section{The \g\ emissions from PSR~J2021$+$3651}

A \g\ pulsar PSR~J2021$+$3651 is spatially coincident with the 4HWC~J2021$+$3650. It is significant to investigate whether PSR~J2021$+$3651 can power 4HWC~J2021+3650. The acceleration of protons in PWN and the subsequent production of hadronic \g\ signals are theoretically feasible \citep{Lemoine2015,Mitchell2026}. For example, mass-loading of the pulsar wind by ions from the surrounding interstellar medium is potential hadronic scenario, which are then accelerated in the wind or at the termination shock \citep{Lyutikov2003,Morlino2015}. However, current evidence suggests that leptonic scenario is the dominant process. For the Crab Nebula, lepto-hadronic models remain speculative \citep{Cao2021b,Aharonian2024}, which constrained by \g\ spectral features.

As efficient electron factories, PWN are powered by energetic pulsars, which inject relativistic electrons and positrons in their magnetosphere. \cite{Wilhelmi2022} explores the potential of young, energetic pulsars to power UHE sources particularly those recently discovered by LHAASO with spectra extending beyond 100 TeV \citep{Cao2021a}. The key factor determining whether a pulsar can accelerate electrons to PeV energies is its spin-down luminosity $\dot{E}$. The magnetic field strength of PWN should be low enough with typically a few tens of $\mu$G to avoid severe synchrotron losses and to allow sufficient conversion of spin-down power into \g\ emitting electrons via IC scattering off the CMB. An absolute maximum photon energy is derived as $E_{\gamma, \rm max} \approx 0.9 \dot{E}_{36}^{0.65}\ \rm PeV$. For most pulsars, this potential-drop limit is more restrictive than synchrotron losses, meaning the pulsar’s rotational power directly sets the upper bound on achievable particle energies. Considering potential association between LHAASO~J2018$+$3651 and PSR~J2021$+$3651, $E_{\gamma, \rm max} \approx 1.99 \ \rm PeV$ and $E_{e, \rm max} \approx 3.69 \ \rm PeV$ are derived in \cite{Wilhelmi2022}. Such energy of accelerated proton is high enough to explain the \g\ origin of 4HWC~J2021$+$3650. However, the offset between LHAASO~J2018$+$3651 and PSR~J2021$+$3651 is 0.42$^{\circ}$. Thus, the connection between LHAASO~J2018$+$3651 and PSR~J2021$+$3651 is still under debate.

A key conditions under which TeV \g\ halos form around pulsars is the energy density of relativistic electrons $\epsilon_e$ relative to that of the interstellar medium $\epsilon_{\rm ISM}$ \citep{Giacinti2020}. Only when $\epsilon_e \lesssim \epsilon_{\rm ISM}$ can the emission be classified as a true halo, where electrons have escaped into the unperturbed ISM. Halos are expected only at late times ($t \gtrsim 100$ kyr), after the pulsar has escaped its SNR and electrons diffuse into the ISM. Using two estimators related to pulsar spin-down properties and TeV \g\ luminosity, it find that most TeV-bright sources have $\epsilon_e \gtrsim \epsilon_{\rm ISM}$ indicating that their emission originates from within the PWN, where the pulsar dynamically dominates. Only when $\epsilon_e \lesssim \epsilon_{\rm ISM}$ (such as Geminga \citep{Abeysekara2017}) can the emission be classified as a true halo, where electrons have escaped into the unperturbed ISM. Considering $\epsilon_{\rm e} = E_{\rm inj}/V$ and $E_{\rm inj} = \dot{E}/\tau_{\rm c}$, we obtain $\epsilon_{\rm e} \approx 89\ \rm eV/cm^3$ for PSR~J2021$+$3651. This indicates that PSR~J2021+3651 is hard to form TeV halo.

\section{Conclusions}

Berkeley~87 is promising source for accelerating particles as other stellar clusters. We report the extended \g\ emissions around the stellar cluster Berkeley~87 using 17 yr of \lat\ data. The \g\ source 4FGL~J2022.6$+$3716 is potentially associated with Berkeley~87. The extended \g\ emissions has an angular extension of $0.36^{\circ}$ which can be modeled by a radial Gaussian profile. The \g\ spectrum follows a PL distribution with a photon index of $2.68 \pm 0.34$. 

The total mass of the molecular cloud and HII region was found to be approximately $21.62 \times 10^2\ M_{\odot}$, with an average target proton density $368\ \rm cm^{-3}$. The extended \g\ emission observed around Berkeley 87 is unlikely to be dominated by PSR~J2021$+$3651 or its associated PWN. Instead, it is better explained by hadronic processes involving CRs accelerated within the star cluster, interacting with ambient gas.

The cluster hosts at least 15 OB stars and one WR star. The total stellar wind kinetic power exceeds $10^{36}\ \rm erg\ s^{-1}$. Even with a conservative CR acceleration efficiency of 0.1\%–1\%, the cluster can supply the required proton energy ($W_{\rm p} \sim 5 \times 10^{49}\ \rm erg$ above 5 GeV). The best-fit proton index $\alpha_{\rm p} = 2.82$ is typical for diffusive shock acceleration at stellar-wind termination shocks and wind-wind collision zones.

The analysis of PSR~J2021$+$3651 shows that its off-pulse emission exhibits a point-source nature, indicating a magnetospheric origin rather than a surrounding PWN or halo. The derived electron energy density $89\ \rm eV/cm^3$ is significantly higher than the typical ISM energy density, suggesting that PSR~J2021$+$3651 is unlikely to form a TeV \g\ halo. The pulsar’s spin-down power could theoretically accelerate particles to PeV energies, which explain the UHE \g\ emission from 4HWC~J2021$+$3650. 

\section*{Data Availability}

The \lat\ data used in this work are publicly available and are provided online at the NASA-GSFC Fermi Science Support Center \footnote{https://fermi.gsfc.nasa.gov/cgi-bin/ssc/LAT/LATDataQuery.cgi}. All the template of gas can be found in SkyView \footnote{https://skyview.gsfc.nasa.gov/current/cgi/query.pl}.

\section*{Acknowledgements}

Ziwei Ou is supported by the National Natural Science Foundation of China (NSFC, Grant No. 12393853). Xiaolong Yang thanks for the support from the National Science Foundation of China (12103076) and the Shanghai Sailing Program (21YF1455300)



\bibliographystyle{mnras}
\bibliography{example} 





\bsp	
\label{lastpage}
\end{document}